# Cation and anion substitutions in hybrid perovskites: solubility limits and phase stabilizing effects

Frederike Lehmann[1,2], Silvia Binet[1], Alexandra Franz[1], Andreas Taubert[2], and Susan Schorr[1,3]

[1]Helmholtz-Zentrum Berlin for Materials and Energy, 14109 Berlin, Germany;
[2]Institute of Chemistry, University Potsdam, 14476 Potsdam, Germany;
[3]Institute of Geological Sciences, Freie Universitaet Berlin, 12249 Berlin, Germany

*Abstract* — Organic or inorganic (A) metal (M) halide (X) perovskites ($AMX_3$) are semiconductor materials setting the basis for the development of highly efficient, low-cost and multijunction solar energy conversion devices. The best efficiencies nowadays are obtained with mixed compositions containing methylammonium, formamidinium, Cs and Rb as well as iodine, bromine and chlorine as anions. The understanding of fundamental properties such as crystal structure and its effect on the band gap, as well as their phase stability is essential. In this systematic study X-ray diffraction and photoluminescense spectroscopy were applied to evaluate structural and optoelectronic properties of hybrid perovskites with mixed compositions.

## I. Introduction

Materials with Perovskite type structure are currently one of the most captivating ideas in the field of solar energy conversion. The efficiency of hybrid perovskite-based devices rose up impressively fast, even faster than the fundamental understanding of the material itself.

Using the hybrid perovskite $CH_3NH_3PbI_3$ ($MAPbI_3$) as absorber materials efficiencies up to 22.1% can be reached [1], but the best efficiencies nowadays are obtained with mixed compositions containing methylammonium (MA) ($CH_3NH_3^+$), formamidinium (FA) ($CH(NH_2)_2^+$), Cs, Rb as well as iodine, bromine and chlorine as the anions.

The $MAPbI_3$ hybrid perovskite is the most studied one amongst these compounds. At room temperature $MAPbI_3$ crystallizes in a tetragonal structure with the space group $I4/mcm$, with a $PbI_6$ - octahedra tilting around [001] [2]. The orientation of the MA molecule in <221> according to the tetragonal unit cell is observed [2] which represents a complex orientation distribution of the molecule in between the $PbI_6$-octahedra.

$MAPbCl_3$ as well as $MAPbBr_3$ crystallize in a cubic structure with the space group $Pm\bar{3}m$ at room temperature [3]. In this structure the MA molecule has no preferred orientation and shows a statistical orientation distribution, i. e. complete disorder.

$CsPbI_3$, $RbPbI_3$ crystallize in a non-perovskite structure at room temperature (orthorhombic structure with space group $Pnma$,) and display a too large band gap for photovoltaic applications [4]. An increased stability of device performances was proved in $Cs_xMA_{(1-x)}PbI_3$, [5] for an optimized amount of cesium.

For $FAPbI_3$ different RT phases are reported: a yellow non-perovskite δ-phase (hexagonal space group $P6_3mc$) and an unstable black perovskite type α-phase where the structure is discussed as cubic (space group $Pm\bar{3}m$) or trigonal (space group $P3m1$) [6].

In the mixed $FA_{1-x}Cs_xPbI_3$ composition a perovskite type structure is energetically more favorable and is stable at room temperature [7]. The same effect has been more recently observed when low amounts of rubidium are inserted in the $FAPbI_3$ structure [8].

The understanding of fundamental properties of these solid solutions such as the crystal structure and its effect on the band gap, as well as their phase stability and solubility limits is essential. Thus we performed a systematic study of hybrid perovskite solid solutions with mixed compositions, including the end members, from a structural, thermal and optical perspective.

## II. Experimental

### A. Synthesis of powder samples

Powder samples of the end members and corresponding mixtures were prepared by the following general route [3]: $APbX_3$ (A=MA, FA, Cs, Rb; X=I, Br, Cl) were obtained by dissolving AX and $PbX_2$ in equimolar amounts in γ-butyrolactone (GBL) and /or N,N-dimethylformamide (DMF). The formed solution was stirred overnight at a certain temperature (T1) and finally evaporated in a petri dish at elevated temperatures (T2). Synthesis parameters are summarized in table 1.

$MAPbCl_3$ and Cl-rich $MAPb(I_{1-x}Cl_x)_3$ was synthesized by a complex procedure described in [9]. A methylamine solution (40 wt% in $H_2O$) in a 100 ml triple neck round bottom flask with dropping funnel was cooled down to 5°C and neutralised with concentrated aqueous hydrochloric acid (HCl, fuming, 37%). To prevent the formation of $PbCl_2$ as a chemical by-product, excess HCl (approx. 2 ml) was additionally added. This solution was heated in an oil bath to 100°C. An aqueous solution of 2.21g lead-(II) acetate trihydrate ($Pb(CH_3COO)_2 \cdot 3H_2O$, 99.5%) and 9ml $H_2O$ was then added dropwise. A fine white precipitate resulted. To avoid hydration of the product at temperatures below 40°C, the reaction mixture was filtered at 80 °C, washed with ethanol

(absolute 99.8%) and afterwards dried at 50°C for 1-2 days in drying cabinet.

*B. Characterization of phase stability, structural and optoelectronic properties*

X-ray powder diffraction data were collected at the KMC-2 beamline at the synchrotron source BESSY II (HZB) using the Diffraction end-station ($\lambda$=1.5406(1) Å). Additional powder diffraction data were measured using both a PANalytical X'Pert PRO MRD diffractometer and a Bruker D8 Advance diffractometer with Bragg-Brentano geometry (Cu K$\alpha$ radiation). The diffraction data were analyzed applying a LeBail refinement procedure.

TABLE I
SUMMARY OF SYNTHESIS PARAMETERS

| compound | solvent | T1 /°C | T2 /°C |
|---|---|---|---|
| MAPbI$_3$ [2,3] | GBL | 60 | 110 |
| MAPbBr$_3$ | DMF | 25 | 80 |
| MAPb(I$_{1-x}$Cl$_x$)$_3$ (I-rich) | GBL, DMF | 60 - 85 | 110 |
| MAPb(I$_{1-x}$Br$_x$)$_3$ (I-rich) | GBL, DMF | 60 | 110 |
| MAPb(1-xBrx)$_3$ (Br-rich) | GBL, DMF | 60 | 80 |
| FAPbI$_3$ [3] | GBL | 60 | 115 |
| FA$_x$MA$_{1-x}$PbI$_3$ | GBL | 60 | 115 |
| CsPbI$_3$ | DMF | 60 | 110 |
| Cs$_x$MA$_{1-x}$PbI$_3$ (Cs-poor) | GBL, DMF | 60 | 120 |
| RbPbI$_3$ | DMF | 60 | 120 |
| Rb$_x$MA$_{1-x}$PbI$_3$ (Rb-poor) | GBL | 60 | 110 |

For photoluminescence (PL) measurements the powder samples were placed on a glass substrate and irradiated with a diode laser at 409 nm as excitation source. Excitation density was adjusted by changing neutral density filters. The emission signal was detected with a thermoelectrically cooled CCD camera. Two parabolic mirrors were used to obtain an excitation spot size of 30 μm and 100 μm. By manual repositioning of the sample several emissions spots were measured for each sample. No inert atmosphere was used, however the sample proved to be stable for the required time of the measurements.

III. RESULTS AND DISCUSSION

*A. MAPb(I$_{1-x}$Cl$_x$)$_3$*

Substituting the iodine anion by chlorine in MAPbI3 improves the air stability of the hybrid perovskite and furthermore, leads to an increased charge transfer, a lower crystal impedance and a lowered recombination rate [11], [12]. With regard to the substitution limits different and partly contradicting values can be found in literature.
A visual inspection of the XRD pattern of the MAPb(I$_{1-x}$Cl$_x$)$_3$ powder samples shows already the presence of a large miscibility gap. The results of the LeBail refinement accompanied by a careful linewidth analysis revealed limits of the miscibility gap in the solid solution system of MAPb(I$_{1-x}$Cl$_x$)$_3$ hybrid perovskites with 3.1(1.1) mol-% MAPbCl$_3$ in MAPbI$_3$ and 1.0(1) mol-% MAPbI$_3$ in MAPbCl$_3$ [13]. It has to be noted that the lattice parameters of the single phase solid solutions determined by LeBail refinement of the powder diffraction data are not following Vegard's law [14]. Thus a simple conclusion of the chlorine or iodine content in the single phase MAPb(I$_{1-x}$Cl$_x$)$_3$ compounds from their lattice parameters leads to wrong values.

Photoluminescence measurements on MAPbI$_3$ single crystals and an iodine-rich MAPb(I$_{1-x}$Cl$_x$)$_3$ solid solution (for details on the crystal growth method see [15]) did not show a significant increase in the band gap energy. Probably there is no effect because of the small substitutional amount and a possible band gap bowing behavior.

*B. MAPb(I$_{1-x}$Br$_x$)$_3$*

The solid solution MAPb(I$_{1-x}$Br$_x$)$_3$ is described in literature [16] to show a complete solubility. Our X-ray diffraction studies on powder samples revealed a miscibility gap in between $0.29 \leq x \leq 0.92$ ($\pm 0.02$), determined by LeBail refinement (see figure 1) of the data accompanied by a careful linewidth analysis.

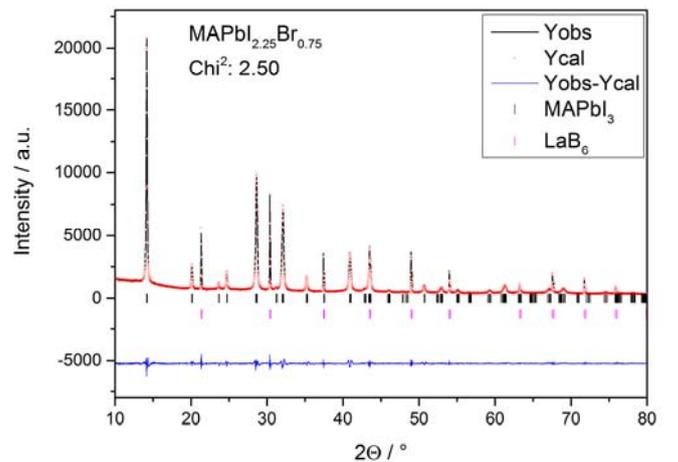

Fig.1. LeBail analysis of powder diffraction data, exemplarily shown for MAPbI$_{2.25}$Br$_{0.75}$. LaB$_6$ was added as internal standard to ensure a precise determination of the lattice parameter.

Nevertheless the single phase regions of iodine-rich and bromine-rich MAPb($I_{1-x}Br_x$)$_3$ solid solutions are much more extended in comparison to the MAPb($I_{1-x}Cl_x$)$_3$ mixture. This finding is reasonable with regard to the ionic radii of $I^-$, $Br^-$ and $Cl^-$. It should be noticed, that a complete solubility was found for the *MAPb($Br_{1-x}Cl_x$)$_3$* series.

Photoluminescence measurements on MAPb($I_{1-x}Br_x$)$_3$ powder samples show a peak shift to higher energies with increasing x value. This can be explained as an increase in band gap energy by substituting iodine for bromine in MAPbI$_3$.

### C. $Cs_xMA_{1-x}PbI_3$

LeBail refinements of the X-ray diffraction data of $Cs_xMA_{1-x}PbI_3$ have shown that in the compounds with x = 0.95 and x = 0.9 no Bragg peaks assigned to MAPbI$_3$ are observed, indicating the complete integration of methylammonium in the orthorhombic lattice of CsPbI$_3$. This is also confirmed by the trend observed in the orthorhombic lattice constant c that increases with the amount of methylammonium. On the other hand, substituting methylammonium by cesium is possible for x<0.15. Thus the miscibility gap in $Cs_xMA_{1-x}PbI_3$ extends in between $0.15 < x \leq 0.9$.

The photoluminescence measurements revealed a nonlinear behavior with the amount of cesium is observed for the $Cs_xMA_{1-x}PbI_3$ series. In fact, for the sample with x=0.05 and 0.07 the emission is red-shifted when compared to that of MAPbI$_3$, whilst it becomes blue shifted for x=0.1 and x=0.15 (see figure 2). In the case of the sample with the smallest amount as possible of cesium, x=0.025, the signal is almost perfectly overlapped to the one of MAPbI$_3$.

### D. $Rb_xMA_{1-x}PbI_3$

In the series including rubidium the lattice parameter determined by LeBail refinement of the X-ray diffraction data do not show a clear dependency as in the other series investigated, The dimensions of the unit cell vary considerably less consistently, due to the low solubility of rubidium (< 3.5 %) which comes from its smaller cationic radius when compared to cesium. This makes the integration in MAPbI3 crystal structure unfavorable.

Even though only small rubidium amounts permitted the formation of single phase solid solutions, at concentrations as low as x = 0.015 and 0.035, the effects of the cation on the optical and thermal properties are already visible and comparable to the ones reported in the case of cesium.
In fact, a red shifted emission was observed for low amounts of guest cations in both series ($Rb_xMA_{1-x}PbI_3$ and $Cs_xMA_{1-x}PbI_3$), precisely with x=0.05 and 0.07 Cs content and x=0.015 and 0.035 Rb content. This result is in contradiction with the general trend in hybrid perovskites discussed in literature, according to which the bandgap increases when the pseudocubic lattice parameter contracts [17].

### E. $FA_{1-x}MA_xPbI_3$

The synthesis of FAPbI$_3$ resulted a black powder which was attributed to the α-phase (perovskite type) according to the X-ray diffraction pattern. The powder degraded within 3 hours to a yellow powder which was attributed to the non-perovskite δ-phase. Bragg peaks of both phases are visible in the X-ray diffraction pattern. Substituting FA by MA shows a stabilizing effect: in the X-ray diffraction pattern of $FA_{1-x}MA_xPbI_3$ with x ≥ 0.2 only Bragg peaks of the cubic perovskite α-phase are visible (see figure 3).

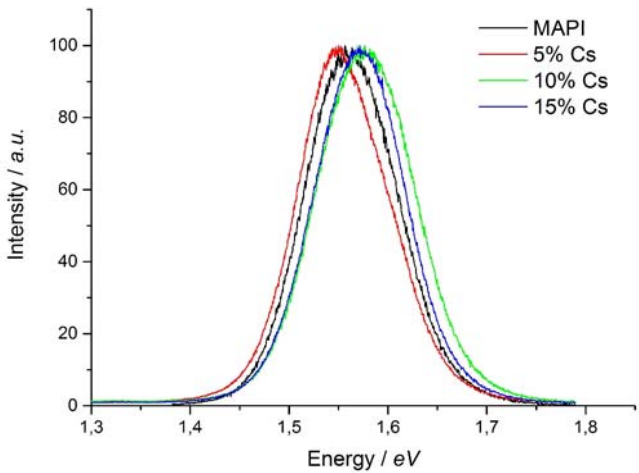

Fig. 2. Photoluminescence spectra of the $Cs_xMA_{1-x}PbI_3$ series with $0 \leq x \geq 0.15$

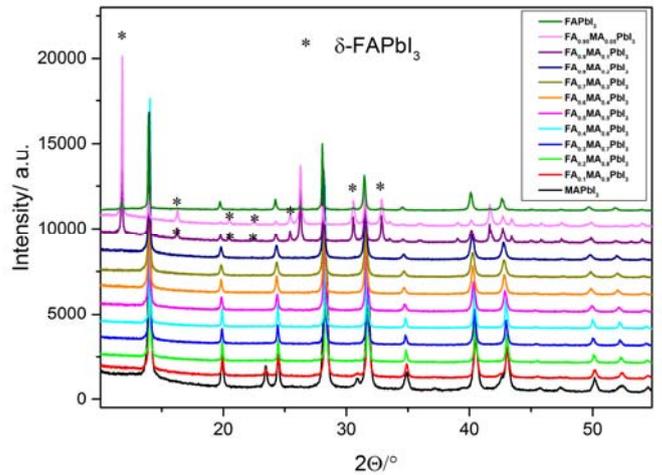

Fig. 3. X-ray powder diffraction pattern of the *$FA_{1-x}MA_xPbI_3$* series. Bragg peaks marked by a star belong to the non-perovskite δ-phase. The other Bragg peaks can be attributed to the cubic or the tetragonal perovskite phase, respectively.

## IV. Conclusions

In solid solution series substituting the "A" cation as well as the anion in MAPbI$_3$ and FAPbI$_3$ miscibility gaps occur in which two phases with the structure of the according end member coexist. The dependency of the lattice parameters from the chemical composition violets partly Vegard's rule. In the case of decreasing lattice parameters within a substitution series a contraction of the crystal structure is obtained. The energy gap of perovskite structured materials is controlled by the orbital overlap of the BX$_6$ octahedra. In fact, changing the size of the anion not only reduces the unit cell size, but also decreases the orbital size of the "X" anion, and consequently the B-X orbital overlap. This does not happen in the case of the substitution of the "A" cation, which is not involved in the band levels: assuming no additional electronic effects, the energy variation of the photoluminescence signal can probably be ascribed to the structural distortions. The lattice contraction and the octahedral rotations have been proposed to be the two competing effects that can explain the non-linear bandgap behavior [18].